\input harvmac
\input epsf
\input ourdefs.defs
\writedefs

\overfullrule=0mm
\hfuzz 10pt

\font\ninerm=cmr9
\font\ninei=cmmi9
\font\nineb=cmbx9

\font\sixi=cmmi6
\def\ninemath{
\textfont0=\ninerm
\textfont1=\ninei \scriptfont1=\sixi \scriptscriptfont1=\scriptfont1}

\def\fig#1#2#3{
\nobreak
\par\begingroup\parindent=0pt\leftskip=1cm\rightskip=1cm\parindent=0pt
\baselineskip=12pt \ninerm \ninemath
\midinsert
\centerline{\hbox{#3}}
\vskip 6pt
{\nineb Fig.~#1:}~#2\par
\vskip 6pt\endinsert
\endgroup\par
\goodbreak
}

\def\({\left(}  \def\){\right)}
\def\[{\left[}  \def\]{\right]}
\def\h{{1\over2}}
\def\th{\textstyle{1\over2}}
\def\ra{\rightarrow}
\def\bra{\langle}       \def\ket{\rangle}

\font\cmss=cmss10 \font\cmsss=cmss10 at 7pt
\def\IZ{\relax\ifmmode\mathchoice
{\hbox{\cmss Z\kern-.4em Z}}{\hbox{\cmss Z\kern-.4em Z}}
{\lower.9pt\hbox{\cmsss Z\kern-.4em Z}}
{\lower1.2pt\hbox{\cmsss Z\kern-.4em Z}}\else{\cmss Z\kern-.4em Z}\fi}

\catcode`\@=11 
\def\eqna#1{\DefWarn#1\wrlabeL{#1$\{\}$}%
\xdef #1##1{(\noexpand\relax\noexpand\checkm@de%
{\s@csym\the\meqno\noexpand\f@rst{##1}}{\hbox{$\secsym\the\meqno##1$}})}
\global\advance\meqno by1}
\catcode`\@=12 

\def\pre#1{({\tt\hbox{#1}})}

\lref\Drouffe{C.~Itzykson, J.--M.~Drouffe, {\sl Statistical field theory},
Cambridge University Press, 1989.}

\lref\SMJ{M.~Sato, T.~Miwa and M.~Jimbo, {\sl Holonomic 
quantum fields}, Publ. RIMS Kyoto Univ. 14 (1978) 223; 15 (1979) 
201, 577, 871; and 16 (1980) 531.}

\lref\BB{O.~Babelon, D.~Bernard, {\sl From form factors to correlation 
functions: the Ising model}, Phys.~Lett. B~288 (1992) 113--120 \pre 
{hep-th/9206003}.}

\lref\BL{D. Bernard, A. Leclair, {\sl Differential equations for 
sine-Gordon correlation functions at the free fermion point}, 
Nucl.~Phys.~B 426 (1994) 534--558; Erratum-ibid. B 498 (1997) 619--621
\pre{hep-th/9402144}.}

\lref\Karowski{B. Berg, M. Karowski, P. Weisz, {\sl Construction of
Green's functions from an exact $S$ matrix}, Phys. Rev. D 19 (1979)
2477--2479.}

\lref\LeClair{A.~LeClair, {\sl Spectrum generating affine Lie algebras in
massive field theory}, Nucl.~Phys. B~415 (1994) 734--780
\pre{hep-th/9305110}.}

\lref\Luther{A.~Luther, I.~Peschel, {\sl Calculation of critical exponents 
in two dimensions from quantum field theory in one dimension}, 
Phys.~Rev.~B 12 (1975) 3908--3917.}

\lref\McCoy{T. T. Wu, B. M. McCoy, C. A. Tracy, E. Barouch, {\sl
Spin-spin correlation functions for the two-dimensional Ising model:
exact theory in the scaling region}, Phys.~Rev. B~13 (1976) 316--374.}

\lref\mccoywuI{B. M. McCoy, T. T. Wu, {\sl 
Two-dimensional Ising field theory in a magnetic field: breakup of the cut 
in the two-point function}, Phys.~Rev. D~18 (1978) 1259--1267.}

\lref\mccoywuII{E. Barouch, B. M. McCoy, T. T. Wu, {\sl 
Zero-field susceptibility of the two-dimensional Ising model near
$T_c$}, Phys. Rev. Lett. 31 (1973) 1409--1411.}

\lref\Perk{W. P. Orrick, B. Nickel, A. J. Guttmann, J. H. H. Perk, {\sl 
The susceptibility of the square lattice Ising model: new developments}, 
J. Statist. Phys. 102 (2001) 795--841 \pre{cond-mat/0103074}.}

\lref\us{P.~Fonseca, A.~Zamolodchikov, {\sl Ising field theory in a 
magnetic field: analytic properties of the free energy}, J.~Stat.~Phys. 
110 (2003) 527--590 \pre {hep-th/0112167}.}

\lref\McCoybook{T. T. Wu, B. M. McCoy, {\sl The two-dimensional ising 
model}, Harvard University Press,~1973.}

\lref\Mussardo{G. Delfino, G. Mussardo, P. Simonetti, {\sl Non-integrable
quantum field theories as perturbations of certain integrable models},
Nucl. Phys. B 473 (1996) 469--508 \pre{hep-th/9603011}.}

\lref\tracy{J.~Palmer, M.~Beatty, C.~A.~Tracy, {\sl Tau functions for the
Dirac operator on the Poincar\'e disk}, Commun.~Math.~Phys. 165 (1994)
97--174 \pre{hep-th/9309017}.}

\lref\youguys{B.~Doyon, P.~Fonseca, in preparation.}

\lref\KadanoffCeva{L. P. Kadanoff, H. Ceva, {\sl Determination of an
operator algebra for the two-dimensional Ising model},
Phys. Rev. B 3 (1971) 3918--3939.}

\lref\saleur{A. Leclair, F. Lesage, S. Sachdev, H. Saleur, {\sl 
Finite temperature correlations in the one-dimensional quantum Ising 
model}, Nucl. Phys. B 482 (1996) 579--612 \pre{cond-mat/9606104}.}

\lref\MW{B. M. McCoy, J. H. H. Perk, T. T. Wu, {\sl Ising field theory: 
quadratic difference equations for the $n$-point Green's functions on the 
lattice}, Phys. Rev. Lett. 46 (1981) 757--760.}

\lref\PerkI{J. H. H. Perk, {\sl Equations of motion for the transverse 
correlations of the one-dimensional $XY$-model at finite temperature}, 
Phys. Lett. A 79 (1980) 1--2.}

\lref\PerkIII{J. H. H. Perk, {\sl Quadratic identities for the Ising model 
correlations}, Phys. Lett. A 79 (1980) 3--5.}

\Title{
\vbox{\baselineskip12pt\hbox{hep-th/0309228}\hbox{RUNHETC-2003-28}}
}
{
\vbox{
\centerline{Ward Identities}{\centerline{}}{\centerline{and
Integrable Differential Equations}}{\centerline{}}{\centerline{in
the Ising Field Theory}} }}
\medskip
\centerline{P.~Fonseca and A.~Zamolodchikov} \medskip
\centerline{\sl Department of Physics and Astronomy, Rutgers University}
\centerline{\sl Piscataway, NJ 08855-0849, USA}

\vskip 0.6in

We show that the celebrated Painlev\'e equations for the Ising
correlation functions follow in a simple way from the Ward
Identities associated with local Integrals of Motion of the
doubled Ising field theory. We use these Ward Identities to derive
the equations determining the matrix elements of the product
$\sigma(x)\sigma(x')$ between any particle states. The result is
then applied in evaluating the leading mass corrections in the Ising
field theory perturbed by an external magnetic field.

\Date{9/2003}

\newsec{Introduction}

It is well known since the pioneering work of Barouch, Tracy, Wu and McCoy
\McCoy\ that the spin-spin correlation function of the Ising model at zero
external field is expressed in terms of suitable solution of certain
ordinary differential equation, the so-called Painlev\'e III equation. 
Later this result was rederived and generalized via different
approaches \refs{\SMJ, \BB, \BL}. While Sato, Miwa and Jimbo 
\SMJ\ used the theory of monodromy preserving deformations of ordinary 
differential equations, the derivation of Babelon and Bernard \BB\ was 
based on exact form-factors \Karowski. One of the goals of this paper is to 
present yet another derivation of this classic result, which we believe is 
somewhat simpler and more straightforward from the field-theoretical 
point of view. It is based on the well known 
fact that the Ising field theory, being a theory of free fermions, 
possesses an infinite set of local integrals of motion. The differential 
equation for the spin-spin correlation function follows directly from the 
associated Ward identities. The simplest way of derivation along these 
lines utilizes the local integrals of motion of the ``doubled'' Ising field 
theory, which consists of two identical copies of the Ising field theory, 
with no interaction between the copies. Various advantages of such 
``doubling'' were recognized before, see e.g. \refs{\Luther, \Drouffe}. The
doubled Ising field theory is equivalent to free Dirac fermion theory, and
it is known to have $\widehat{SL(2)}$ Kac-Moody symmetry generated by local
charges \LeClair. The derivation based on the corresponding Ward 
identities is presented in Section~4 below.

The local integrals of motion act on particle states in a simple way,
and therefore the approach based on the Ward identities applies as
well to all matrix elements of the form
\eqn\bilocal{
\bra\, A(\beta_{1}') \ldots A(\beta_{M}') \mid\, \sigma(x)
\sigma(x')\, \mid A(\beta_{1}) \ldots A(\beta_{N})\,\ket~,
}
where $A(\beta_i)$ stand for the Ising free fermions with rapidities
$\beta_i$, and $\sigma(x)$ denotes the spin field. We derive the
equations determining these matrix elements in Section 4. 

Our particular interest in the matrix elements \bilocal\ was
motivated by the applications to the Ising field theory with nonzero
external field $h$. Adding the term $h\,\int\,\sigma(x)\,d^2 x$ to the 
free-fermion action gives rise to an interesting interacting 
field theory \mccoywuI\ which describes the scaling domain of the Ising 
model in a magnetic field. Generally, this field theory is not integrable, 
hence no exact solution is available. However, at small field, $h << 
m^{15/8}$, perturbation theory in $h$ can produce useful results. As usual, 
such perturbative calculations involve matrix elements of the type
\bilocal. For instance, perturbative calculation of the vacuum
energy requires the correlation functions, and numerical solution of
the corresponding exact differential equation yields the coefficients
of the $h^2$ expansion with high precision \refs{\mccoywuII, \Perk}. The 
matrix elements \bilocal\ involving particles are used in the evaluation 
of the perturbative corrections to particle masses and
scattering amplitudes. In particular, the $\sim h^2$ correction to the 
mass is obtained by evaluating the self-energy part, i.e.
\eqn\selfenergy{
\delta m^2 = - h^2\,\int\,\bra\, A(\beta)\mid\,\sigma(x)
\sigma(0)\,\mid A(\beta) \,\ket_{\rm irred}~d^2 x ~,
} 
where $\bra \ldots \ket_{\rm irred}$ denotes the connected one-particle
irreducible matrix element, and the integration (which makes the
result independent of $\beta$) is taken over the Euclidean
plane I$\!$R$^2$. The integral here can be computed numerically,
using exact differential equation determining the matrix element
involved. The results were previously quoted in \us\ without
derivation. Here, in Section 5 below, we fill this gap by
presenting some details of this calculation.

\newsec{The Ising Field Theory}

In this section we describe the structure of the Ising field theory in
some detail. This is done mostly in order to fix our notations and
conventions; the reader familiar with the subject can safely skip this
Section (except maybe for the important Eqs.~\asigmaIII, \asigmaIV).

As is well known (see e.g. \Drouffe), the Ising Field Theory with 
zero magnetic field is a free Majorana fermion theory, described by the 
standard euclidean action
\eqn\action{
{\cal A}_{\rm FF} = {1\over2\pi}\int
\[\psi\bar\partial\psi + \bar\psi\partial\bar\psi + im\,\bar\psi\psi\]
\,d^2x~. 
} 
Here we have assumed that the theory is defined on an
infinite plane I$\!$R$^2$, whose points $x$ are labelled by
cartesian coordinates $({\rm x},{\rm y})=({\rm x}(x), {\rm
y}(x))$, and $d^2x \equiv d{\rm x}\,d {\rm y}$; In what follows we
typically interpret ${\rm y}$ as the euclidean time direction.
Complex coordinates are defined as ${\rm z}(x) = {\rm x}+i{\rm
y}$, $\bar{\rm z}(x) = {\rm x}-i{\rm y}$, and the derivatives
$\partial$, $\bar
\partial$ in \action\ stand for $\partial_{\rm z} =
\h(\partial_{\rm x}-i\partial_{\rm y})$ and $\partial_{\bar {\rm
z}} = \h(\partial_{\rm x}+i\partial_{\rm y})$, respectively.
The chiral components $\psi(x)$, $\bar\psi(x)$ of the Majorana fermi field
obey the linear field equations
\eqn\eom{
\bar\partial\psi = {im\over2}\,\bar\psi~,
\indent
\partial\bar\psi = -{im\over2}\,\psi~,
}
and their normalization in the action \action\ corresponds to the
following short-distance limit of the two-point operator products
\eqn\normpsi{
({\rm z}-{\rm z}')\,\psi(x)\psi(x')\ra1~,
\quad
(\bar{\rm z}-\bar{\rm z}')\,\bar\psi(x)\bar\psi(x')\ra1~,
\indent{\rm as }\quad |x-x'|\ra0~.
}

Typical observables of the Ising field theory are described in terms of the 
order and disorder fields, $\sigma(x)$ and $\mu(x)$ respectively 
\KadanoffCeva. These fields are non-local with respect to the fermi 
fields; the operator products 
\eqn\product{ 
\psi(x)\sigma(x_0)~,\indent
\bar\psi(x)\sigma(x_0)~,\indent 
\psi(x)\mu(x_0)~,\indent
\bar\psi(x)\mu(x_0)~, 
} 
acquire a minus sign when the point $x$ is taken around $x_0$. This 
property does not define the fields $\sigma$ and $\mu$ uniquely, but a 
precise definition can be given in terms of the ``radial quantization'' of 
the theory \action. Since the products \product\ as functions of $x$ obey 
the field equations \eom, it is useful to introduce a complete set of 
solutions, 
\eqn\radialsol{\eqalign{ 
\pmatrix{\eqalign{u_n(x) 
\cr 
\bar
u_n(x)}} &=\(m\over 2\)^{\h-n}\Gamma\(n+\th\) \pmatrix{\eqalign{
e^{i(n-\h)\theta}\,I_{n-\h}(m\,r) 
\cr
-i\,e^{i(n+\h)\theta}\,I_{n+\h}(m\,r)}}~,
\cr
\pmatrix{\eqalign{v_n(x) \cr \bar v_n(x)}} 
&=\(m\over2\)^{\h-n}\Gamma\(n+\th\) 
\pmatrix{\eqalign{
i\,e^{-i(n+\h)\theta}\,I_{n+\h}(m\,r) 
\cr
e^{-i(n-\h)\theta}\,I_{n-\h}(m\,r)}}~,
}}
(here $n\in\IZ$; $r$, $\theta$ are polar coordinates around $x_0$, i.e.
${\rm z}-{\rm z_0} = r\,e^{i\theta}$, $\bar{\rm z}-\bar{\rm z}_0 =
r\,e^{-i\theta}$; $I_\nu$~are modified Bessel functions) and
write the fields $\psi(x)$ and $\bar\psi(x)$ in the products
\product\ as 
\eqn\radial{\eqalign{ 
\psi(x) &=
\sum_{n=-\infty}^\infty a_n u_{-n}(x) +\bar a_n v_{-n}(x)~,
\cr 
\bar\psi(x) &= \sum_{n=-\infty}^\infty a_n \bar
u_{-n}(x) + \bar a_n \bar v_{-n}(x)~.
}}
The coefficients $a_n$ and $\bar a_n$ here are understood as operators
acting on the space of fields; they can be easily shown to satisfy
the canonical anticommutation relations 
\eqn\modes{ 
\{a_n,a_{n'}\} = \delta_{n+n',0}~, 
\indent 
\{\bar a_n, \bar a_{n'}\} = \delta_{n+n',0}~, 
\indent 
\{a_n,\bar a_{n'}\} = 0~. 
} 
The fields $\sigma$ and $\mu$ are defined as the ``primary'' fields with
respect to the above algebra~\modes. In particular, they satisfy
\eqn\asigmaI{ 
a_n\,\sigma(x) = 0~, 
\indent 
\bar a_n\,\sigma(x) = 0~, 
\indent 
a_n\,\mu(x) = 0~, \indent 
\bar a_n\,\mu(x) = 0 
}
for $n>0$, as well as 
\eqn\asigmaII{ 
\eqalign{ a_0\,\sigma(x) &= {\omega\over\sqrt2}\,\mu(x)~, 
\cr \bar a_0\,\sigma(x) &= {\bar\omega\over\sqrt2}\,\mu(x)~, 
}\indent 
\eqalign{ a_0\,\mu(x) &= {\bar\omega\over\sqrt2}\,\sigma(x)~, 
\cr 
\bar a_0\,\mu(x) &= {\omega\over\sqrt2}\,\sigma(x)~, 
}} 
where $\omega = e^{i\pi/4}$ and $\bar\omega = e^{-i\pi/4}$. 
These equations do not determine the normalization of the fields $\sigma$
and $\mu$; we fix it through the short-distance limit of their
operator products
\eqn\normsigma{ 
|x-x'|^{1/4}~\sigma(x)\sigma(x')\ra 1~, 
\quad
|x-x'|^{1/4}~\mu(x)\mu(x')\ra 1~,
\indent\hbox{as}\quad|x-x'|\ra0~. 
}

Applying the operators $a_n$ and $\bar a_n$ with negative $n$ to the 
``primary'' fields $\sigma$ and $\mu$ creates an infinite tower of 
``descendent'' order and disorder fields. It will become important below 
that the lowest of these ``descendents'' are expressed in terms of the 
coordinate derivatives of the ``primary'' order and disorder fields 
themselves. Indeed, for any local field ${\cal O}(x_0)$, its derivatives 
$\partial_{{\rm z}_0} {\cal O}(x_0)$ and $\partial_{\bar{\rm z}_0} {\cal 
O}(x_0)$ can be expressed as
\eqn\derI{\eqalign{
\partial_{{\rm z}_0} {\cal O}(x_0)
=~&{1\over2\pi i}\oint_{x_0}
\[T(x)\,d{\rm z}(x) + \Theta(x)\,d\bar {\rm z}(x)\]{\cal O}(x_0)~,
\cr
\partial_{\bar{\rm z}_0} {\cal O}(x_0)
= -&{1\over2\pi i}\oint_{x_0}
\[\bar T(x)\,d\bar{\rm z}(x) + \Theta(x)\,d{\rm z}(x)\]{\cal O}(x_0)~,
}}
where the contour integrals in the $x$-plane are taken around the point
$x_0$ in counter-clockwise direction and
\eqn\energytensor{
T = -\h\psi\partial\psi~,
\indent
\Theta = -{im\over4}\bar\psi\psi~,
\indent
\bar T = -\h\bar\psi\bar\partial\bar\psi~,
}
are related to the components of the energy-momentum tensor and normalized
according to the usual conformal field theory convention,
$T = -2\pi\, T_{{\rm z}{\rm z}}$, $\Theta = 2\pi\, T_{{\rm z}\bar{\rm
z}}$ and $\bar T = -2\pi\, T_{\bar{\rm z}\bar{\rm z}}$. Of course, the
contour integrals \derI\ are nothing else but the commutators with the 
energy-momentum operators ${\bf P}$ and ${\bf \bar P}$,
\eqn\derII{
\partial_{{\rm z}_0} {\cal O}(x_0)
= -i\,[{\bf P}, {\cal O}(x_0)]~,
\indent
\partial_{\bar{\rm z}_0} {\cal O}(x_0)
= i\,[{\bf \bar P}, {\cal O}(x_0)]~,
}
where
\eqn\Pcharge{
{\bf P} = -{1\over2\pi}\int_{-\infty}^\infty
\[T(x) + \Theta(x)\]\,d{\rm x}~,
\indent
{\bf \bar P} = -{1\over2\pi}\int_{-\infty}^\infty
\[\bar T(x) + \Theta(x)\]\,d{\rm x}~,
}
with the integration taken along an ``equal-time'' slice ${\rm y}
= {\rm constant}$.
If the local field ${\cal O}$ is either $\sigma$ or $\mu$ or one of their
descendents, it is straightforward to express these commutators in terms of
the mode operators $a_n$, $\bar a_n$ in \radial,
\eqn\derIII{\eqalign{
\partial_{{\rm z}_0} {\cal O}(x_0)
&= \h\sum_{n=0}^\infty
\[(2n+1)a_{-n-1}a_n + {m^2\over 2n+1}\bar a_{-n}\bar a_{n+1}\]
{\cal O}(x_0)~,
\cr
\partial_{\bar{\rm z}_0} {\cal O}(x_0)
&= \h\sum_{n=0}^\infty
\[(2n+1)\bar a_{-n-1}\bar a_n + {m^2\over 2n+1} a_{-n}a_{n+1}\]
{\cal O}(x_0)~.
}}
Using the defining equations for $\sigma$ and $\mu$, \asigmaI\ and 
\asigmaII, the above equations \derIII\ yield the lowest descendents
\eqn\asigmaIII{
\eqalign{ a_{-1}\,\sigma(x) &= {\omega\over\sqrt2}\,4\partial\mu(x)~, 
\cr 
\bar a_{-1}\,\sigma(z,\bar z) &= {\bar\omega\over\sqrt2}\,4\partial\mu(x)~, 
} \indent 
\eqalign{
a_{-1}\,\mu(x) &= {\bar\omega\over\sqrt2}\,4\partial\sigma(x)~,
\cr 
\bar a_{-1}\,\mu(x) &= {\omega\over\sqrt2}\,4\partial\sigma(x)~, 
}} 
and
\eqn\asigmaIV{
\eqalign{ a_{-2}\,\sigma(x) &= 
{\omega\over\sqrt2}\,{8\over3}\partial^2\mu(x)~, 
\cr 
\bar a_{-2}\,\sigma(x) &=
{\bar\omega\over\sqrt2}\,{8\over3}\partial^2\mu(x)~, } \indent
\eqalign{ a_{-2}\,\mu(x) &=
{\bar\omega\over\sqrt2}\,{8\over3}\partial^2\sigma(x)~, 
\cr 
\bar a_{-2}\,\mu(x) &=
{\omega\over\sqrt2}\,{8\over3}\partial^2\sigma(x)~. 
}} 

Of course, the fields $\sigma(x)$ and $\mu(x)$ are not mutually local;
again, the product $\sigma(x)\mu(x')$ picks up a minus sign when $x$
is brought around $x'$. In fact, this product admits operator product
expansion which involves the Majorana fields $\psi, {\bar\psi}$ and
local composite fields built from them. Thus, the leading short-distance 
terms are \KadanoffCeva
\eqn\spinmu{ 
\sigma(x)\mu(x') \sim {1\over\sqrt2|x-x'|^{1/4}} 
\Big[ \omega\,({\rm z}-{\rm z'})^{1/2}\,\psi(x') 
+ \bar\omega\,({\rm \bar z} - {\rm \bar z'})^{1/2}\,\bar\psi(x') \Big] 
} 
as $|x-x'|\ra0$.

The Majorana theory \action\ describes the critical behaviour of the
Ising model in the scaling regime near its phase transition point
$T_c$. The theory applies to both phases, just below and just above
the $T_c$, depending on the choice of the sign of the mass parameter $m$
in \action. Our definition in \asigmaII\ corresponds to the
identification of the case $m >0$ with the ordered phase $T<T_c$,
while the case $m<0$ is identified with the disordered phase
$T>T_c$. Of course, the theories at the oposite signs of $m$ are related
to each other through the well-known Kramers-Wannier duality $(\psi,
{\bar\psi}) \leftrightarrow (\psi, -{\bar\psi}), \ \sigma
\leftrightarrow \mu$, and there is no need to consider them
separately. For this reason in what follows we assume that $m$ is
positive, i.e. we interpret our results in terms of the ordered
phase. In this case it is the order field $\sigma$ which develops 
nonzero vacuum expectation value \McCoybook,
\eqn\order{
\bra \,\sigma(x) \,\ket 
\equiv {\bar\sigma} = m^{1/8}\,{\bar s}\,, \indent {\bar s} = 
2^{1/12} e^{-1/8} A^{3/2}}
($A=1.28243\ldots$ is Glaisher's constant), while the disorder field
$\mu$ does not, $\bra \,\mu(x)\,\ket =0$.

Also, \action\ is a free field theory, and of course its Hilbert space
of states ${\cal H}$ associated with
an ``equal-time'' slice ${\rm y}={\rm constant}$ is the Fock space
of free fermions. It is generated by canonical fermionic creation
and annihilation operators, ${\bf a}^\dagger(\beta)$ and ${\bf
a}(\beta)$, subject to the canonical anticommutation relations
\eqn\anti{ 
\{{\bf a}^\dagger(\beta), {\bf a}(\beta')\} 
= 2\pi\,\delta(\beta-\beta')~, \indent 
\{{\bf a}^\dagger(\beta), {\bf a}^\dagger(\beta')\} = 0~, \indent 
\{{\bf a}(\beta), {\bf a}(\beta')\} = 0~. 
} 
Here $\beta$ is the usual rapidity variable
describing the energy-momentum state of the particle, $(p^0, p^1)
= (m\cosh\beta, m\sinh\beta)$. The canonical operators are the
Fourier modes of the Heisenberg field operators $\psi(x)$ and
${\bar\psi}(x)$, i.e. 
\eqn\planewaves{\eqalign{ 
\psi(x) &= \sqrt{m\over2}\int_{-\infty}^\infty {d\beta\over\sqrt{2\pi}}
\,e^{\beta/2} \Big[{\bf a}^\dagger(\beta)\, 
e^{{\rm y}\,m\cosh\beta -i{\rm x}\,m\sinh\beta} - i\,{\bf a}(\beta)\,
e^{-{\rm y}\,m\cosh\beta + i{\rm x}\,m\sinh\beta}\Big]~, 
\cr 
\bar\psi(x) &= \sqrt{m\over2}\int_{-\infty}^\infty {d\beta\over\sqrt{2\pi}}
\,e^{-\beta/2} \Big[{\bf a}^\dagger(\beta)\, 
^{{\rm y}\,m\cosh\beta -i{\rm x}\,m\sinh\beta} + i\,{\bf a}(\beta)\,
e^{-{\rm y}\,m\cosh\beta + i{\rm x}\,m\sinh\beta}\Big]~, 
}}
Below we use the notation 
\eqn\state{
\mid A(\beta_1)\ldots A(\beta_N)\,\ket 
= {\bf a}^\dagger(\beta_1)\ldots {\bf a}^\dagger(\beta_N)\mid 0\,\ket~ 
} 
for an $N$-particle state.

\newsec{Conserved Charges of the Doubled Ising Field Theory}

It was observed in the past that many aspects of the Ising field
theory simplify if one considers a system of two identical copies
of the Ising model, with no interaction between the copies - the
``doubled'' Ising field theory \refs{\Luther, \Drouffe}. Here we also use 
this trick, introducing two copies henceforth referred to as ``copy a'' 
and ``copy b''. We will use the subscript $a$ or $b$ to distinguish
between the fields belonging to the corresponding copy; thus,
$\psi_a (x)\,,\, {\bar\psi}_a (x)\,,\, \sigma_a (x)\,,\, \mu_a (x)$ will
stand for the fermionic and the order-disorder fields from the copy
a, and $\psi_b (x)\,,\, \bar\psi_b (x)\,,\, \sigma_b (x)\,,\, \mu_b
(x)$ will denote these fields from the copy~b. The correlation
functions of the doubled theory factorize in terms of the
correlation functions of the individual copies, in an obvious
manner.

The above straightforward definition has an inconvenient feature
that the fermi fields belonging to different copies commute. This
can be fixed by introducing, in addition to the two Ising copies,
two auxiliary variables, the so-called ``Klein factors'', $\eta_a$
and $\eta_b$. These are assumed to commute with all the observables from 
the original Ising copies, and to satisfy the following defining relations 
\eqn\klein{
\eta_a^2 = 1~, \indent
\eta_b^2 = 1~, \indent 
\eta_a\eta_b = -\eta_b\eta_a~, 
}
and
\eqn\kleinvev{
\bra \eta_a \ket =0~,
\indent
\bra \eta_b \ket =0~.
}
We then modify the definitions of the fields of the doubled Ising
field theory as follows, 
\eqn\fermid{
\eqalign{ 
&\psi_a(x) \to \eta_a\,\psi_a(x)~, 
\indent 
\psi_b(x) \to \eta_b\,\psi_b(x)~, 
\cr
&\bar\psi_a(x) \to \eta_a\,\bar\psi_a(x)~, 
\indent 
\bar\psi_b(x) \to \eta_b\,\bar\psi_b(x)~, 
}} 
and 
\eqn\orderd{
\eqalign{
&\sigma_a(x) \to \sigma_a(x)~,
\cr
&\mu_a(x) \to \eta_a\,\mu_a (x)~,
}\indent
\eqalign{
&\sigma_b(x) \to \sigma_b(x)~,
\cr
&\mu_b(x) \to \eta_b\,\mu_b (x)~.
}}

This modification has no effect on any of the relations in Section 2, 
and hence it does not bring any change to the correlation functions 
involving only the fields from a single copy, either $a$ or $b$. On 
the other hand, the Klein factors give rise to additional sign factors when
factorizing generic correlation functions with entries from both
copies present. For example, $\bra \sigma_a (x) \sigma_b (x)
\,\sigma_a (x')\sigma_b (x')\ket = \bra \sigma(x)\,\sigma(x')\ket^2$, but 
$\bra \mu_a (x) \mu_b (x) \,\mu_a (x')\mu_b (x')\ket = - 
\bra \mu(x)\,\mu(x')\ket^2$, where the expectation values in
the right-hand sides are that of a single Ising field theory.
The desirable effect of this modification is that fermi fields from 
different copies now anticommute inside the correlation functions, 
\hbox{$\psi_a (x)\psi_b (x') = - \psi_b (x')\psi_a (x)$}, etc. Of course, 
with this definition the doubled Ising field theory is identical to the 
theory of a free Dirac fermi field 
\eqn\diracf{
\Psi(x) = \big(\psi_a (x) + i\,\psi_b (x),\ 
{\bar\psi}_a (x) + i\,{\bar\psi}_b(x)\big)~.
}

\indent

The space of states of the doubled Ising field theory involves two
species of free fermi particles, {\it A} and {\it B}, associated
with the copies {\it a} and {\it b}, respectively. The
multiparticle states are generated by the corresponding creation
and annihilation operators ${\bf a}^{\dagger}(\beta)\,, {\bf
a}(\beta)$ and ${\bf b}^{\dagger}(\beta)\,, {\bf b}(\beta)$:
\eqn\statedouble{ 
|A(\beta_1)\ldots A(\beta_N)
B(\beta_{N+1})\ldots B(\beta_M)\ket 
= {\bf a}^\dagger(\beta_1)\ldots {\bf a}^\dagger(\beta_N) 
{\bf b}^\dagger(\beta_{N+1})\ldots {\bf b}^\dagger(\beta_M)|0\ket~. 
}
The operators ${\bf a}^{\dagger}(\beta)\,, {\bf a}(\beta)$ and
${\bf b}^{\dagger}(\beta)\,, {\bf b}(\beta)$ are related to the
fields $\psi_a (x), {\bar\psi}_a(x)$ and $\psi_b (x),
{\bar\psi}_b(x)$, respectively, through equations identical to
Eqs. \planewaves. Of course, each pair ${\bf 
a}^{\dagger}(\beta)\,, {\bf a}(\beta)$ and ${\bf b}^{\dagger}(\beta)\,, 
{\bf b}(\beta)$ obeys the canonical anticommutators \anti, and ${\bf 
a}^{\dagger}(\beta)\,, {\bf a}(\beta)$ anticommute with ${\bf 
b}^{\dagger}(\beta)\,, {\bf b}(\beta)$.

Being a free fermion theory, the doubled Ising field theory
exhibits an infinite set of local Integrals of Motion (IM). We
have no need to describe the whole set here. Certain subset of
these IM forms the commutator algebra $\widehat {SL(2)}$ of the level zero; 
its detailed description is given in Ref.~\LeClair. Here we display only 
the fundamental generating elements of this subset, since these basic IM 
are directly used in the calculations below.

The simplest IM is the $U(1)$ charge associated with phase rotations of 
the Dirac field~\diracf, 
\eqn\Zzerocharge{
{\bf Z}_0 = {1\over2\pi}\int_{-\infty}^\infty
\[\psi_a\psi_b - \bar\psi_a\bar\psi_b\]\, d{\rm x}~.
} 
Also, the energy-momenta $\({\bf P}, {\bar{\bf P}}\)$ associated
with each copy are conserved separately, hence their linear combinations
\eqn\Xcharge{
{\bf X}_1 = {\bf P}_a - {\bf P}_b~,
\indent
{\bf X}_{-1} = {\bf \bar P}_a - {\bf \bar P}_b~,
} 
are conserved as well. Less trivially, the
following integrals are conserved, 
\eqn\Ycharge{\eqalign{ 
{\bf Y}_1 &= {1\over2\pi}\int_{-\infty}^\infty
\[\psi_a\partial\psi_b  + i{m\over2}\bar\psi_a\psi_b\]\,d{\rm x}~,
\cr 
{\bf Y}_{-1} &= {1\over2\pi}\int_{-\infty}^\infty
\[\bar\psi_a\bar\partial\bar\psi_b
-i{m\over2}\psi_a\bar\psi_b\]\,d{\rm x}~,
}}
as one can check by a simple direct computation. Equally straightforward
calculation yields the basic commutators 
\eqn\algebra{\eqalign{
&[{\bf Z}_0,{\bf X}_1] = 2i\,{\bf Y}_1~, 
\cr 
&[{\bf Z}_0, {\bf X}_{-1}] = 2i\,{\bf Y}_{-1}~, 
\cr 
&[{\bf X}_1,{\bf Y}_{-1}] = 2i\(m\over2\)^2\, {\bf Z}_0~, 
\cr 
&[{\bf X}_1,{\bf X}_{-1}] = 0~,
} \indent 
\eqalign{ &[{\bf Z}_0,{\bf Y}_1] = -2i\,{\bf X}_1~, 
\cr &[{\bf Z}_0, {\bf Y}_{-1}] = -2i\,{\bf X}_{-1}~, 
\cr &[{\bf X}_{-1}, {\bf Y}_1] = 2i\(m\over2\)^2\, {\bf Z}_0~, 
\cr &[{\bf Y}_1,{\bf Y}_{-1}] = 0~, 
}} 
which show that the operators ${\bf Z}_0, {\bf X}_{\pm 1}, {\bf Y}_{\pm 1}$ 
generate the affine Lie algebra $\widehat{SL(2)}$ of the level zero 
\foot{The standard Chevalley generators $\{{\bf E}_\pm, {\bf F}_\pm, {\bf 
H}_\pm\}$ for the $\widehat{SL(2)}$ algebra are the linear combinations
$$
{\bf E}_\pm = ({\bf X}_1\pm i\,{\bf Y}_1)/m~, \indent {\bf F}_\pm
= ({\bf X}_{-1}\mp i\,{\bf Y}_{-1})/m~, \indent {\bf H}_\pm = \pm
{\bf Z}_0~.
$$
One can check validity of the Serre relations $[{\bf
E}_\pm,[{\bf E}_\pm,[{\bf E}_\pm,{\bf E}_\mp]]] = 0$ and \hbox{$[{\bf
F}_\pm,[{\bf F}_\pm,[{\bf F}_\pm,{\bf F}_\mp]]] = 0$} by yet
another straightforward computation.}. Further commutators of
these basic generators give rise to other elements of
$\widehat{SL(2)}$, which all have the form of integrals of local
densities, quadratic in terms of the fermi fields and their derivatives
\LeClair. In our calculations below two such elements,
\eqn\algebraII{ 
2i\,{\bf Z}_2 = [{\bf X}_1, {\bf Y}_1]~, \indent
-2i\,{\bf Z}_{-2} = [{\bf X}_{-1}, {\bf Y}_{-1}]~,
} 
will be particularly useful. Explicitly, 
\eqn\Ztwocharge{\eqalign{
{\bf Z}_2 &= {1\over2\pi}\int_{-\infty}^\infty
\[\partial\psi_a\partial\psi_b + \(m\over2\)^2\psi_a\psi_b\]\,d{\rm x}~,
\cr 
{\bf Z}_{-2} &= {1\over2\pi}\int_{-\infty}^\infty
\[\bar\partial\bar\psi_a\bar\partial\bar\psi_b
+ \(m\over2\)^2\bar\psi_a\bar\psi_b\]\,d{\rm x}~.
}}

In the next Section we study the Ward identities associated
with this $\widehat{SL(2)}$ symmetry of the doubled Ising field
theory. Derivations of the Ward identities involve commutators of
the symmetry generators with local fields. Specifically, we will
be interested in the commutators of the generators with the
products of the order or the disorder fields from different
copies, i.e. with the fields $\sigma_a (x) \sigma_b(x)$, $\sigma_a
(x)\mu_b(x)$, $\mu_a (x)\sigma_b(x)$ and $\mu_a(x)\mu_b(x)$. Note
that the densities of the IM \Zzerocharge--\Ycharge,
\Ztwocharge\ are local with respect to such product fields, and
therefore these commutators are again local fields. Of course, the
commutators with ${\bf X}_{\pm 1}$ can be written down immediately
from the definitions~\Xcharge, 
\eqn\Xaction{\eqalign{ 
[{\bf X}_1,{\cal O}_a(x){\cal O}_b(x)] 
&= i\,\partial{\cal O}_a(x){\cal O}_b(x) - i\,{\cal O}_a(x)\partial{\cal 
O}_b(x)~, 
\cr 
[{\bf X}_{-1},{\cal O}_a(x){\cal O}_b(x)] &= -i\,\bar\partial{\cal
O}_a(x){\cal O}_b(x) + i\,{\cal O}_a(x)\bar\partial{\cal O}_b(x)~.
}} 
Derivation of the other commutators are less elementary, but
yet straightforward. One can use the decomposition \radial\ for each of 
the fields $\psi_a$ and $\psi_b$ to express any such  commutator in terms 
of the corresponding radial mode operators $a_n, {\bar a}_n$ and $b_n, 
{\bar b}_n$. For instance,
\eqn\Zzerocomm{
[{\bf Z}_0, {\cal O}_a {\cal O}_b ] = 
-i\(a_0b_0+\bar a_0\bar b_0\){\cal O}_a {\cal O}_b
-i\sum_{n=1}^\infty\(
a_{-n}b_n - b_{-n}a_n + \bar a_{-n}\bar b_n - \bar b_{-n}\bar a_n
\){\cal O}_a {\cal O}_b~.
}
When the primary order or disorder fields are
concerned, Eqs.~\asigmaI, \asigmaII, and \asigmaIII, \asigmaIV\
apply. It is important to keep track of the Klein factors $\eta_a$, 
$\eta_b$ in these calculations to get the signs right. For the commutators 
involving ${\bf Z}_0$ and ${\bf Y}_1$, the results are 
\eqn\Zaction{\eqalign{
&[{\bf Z}_0, \sigma_a(x)\sigma_b(x)] = 0~,
\cr
&[{\bf Z}_0, \sigma_a(x)\mu_b(x)] = -i\,\mu_a(x)\sigma_b(x)~,
}
\indent
\eqalign{
&[{\bf Z}_0, \mu_a(x)\mu_b(x)] = 0~,
\cr
&[{\bf Z}_0, \mu_a(x)\sigma_b(x)] = i\,\sigma_a(x)\mu_b(x)~,
}}
and
\eqn\Yaction{
\eqalign{ [{\bf Y}_1,\sigma_a(x)\sigma_b(x)] &=
-\partial\mu_a(x)\mu_b(x) + \mu_a(x)\partial\mu_b(x)~, \cr [{\bf
Y}_1, \mu_a(x)\mu_b(x)] &= -\partial\sigma_a(x)\sigma_b(x) +
\sigma_a(x)\partial\sigma_b(x)~, \cr [{\bf Y}_1,
\sigma_a(x)\mu_b(x)] &= i\,\partial\mu_a(x)\sigma_b(x) -
i\,\mu_a(x)\partial\sigma_b(x)~, \cr [{\bf Y}_1,
\mu_a(x)\sigma_b(x)] &= -i\,\partial\sigma_a(x)\mu_b(x) +
i\,\sigma_a(x)\partial\mu_b(x)~. }} 
Let us also present two useful commutators involving the operator 
${\bf Z}_2$,
\eqn\Ztwoaction{
\eqalign{ [{\bf Z}_2,\sigma_a(x)\sigma_b(x)] &=
2\partial\mu_a(x)\partial\mu_b(x) - \partial^2\mu_a(x)\mu_b(x) -
\mu_a(x)\partial^2\mu_b(x)~, \cr [{\bf Z}_2,\mu_a(x)\mu_b(x)] &=
2\partial\sigma_a(x)\partial\sigma_b(x)
 - \partial^2\sigma_a(x)\sigma_b(x)
 - \sigma_a(x)\partial^2\sigma_b(x)~.}} 
The operators ${\bf Y}_{-1}$ and ${\bf Z}_{-2}$ satisfy
relations similar to \Yaction\ and \Ztwoaction\ with $\partial$
replaced by $\bar\partial$ and $i$ replaced by $-i$.

In addition to the above commutators, the calculations in Section
4 below will involve the action of the generators on the multiparticle
states $|A(\beta_1)\ldots A(\beta_N)B(\beta_{N+1})\ldots
B(\beta_M)\ket$  of the doubled Ising model. The equations
\eqn\ZactionII{
{\bf Z}_0 = -i\int_{-\infty}^\infty
\Big[{\bf a}^\dagger(\beta){\bf b}(\beta)
- {\bf b}^\dagger(\beta){\bf a}(\beta)\Big] {d\beta\over2\pi}~,
}
and
\eqn\YactionII{\eqalign{
{\bf Y}_1 &= {m\over2}\int_{-\infty}^\infty
e^\beta\Big[
{\bf a}^\dagger(\beta){\bf b}(\beta) +
{\bf b}^\dagger(\beta){\bf a}(\beta)\Big] {d\beta\over2\pi}~,
\cr
{\bf Y}_{-1} &= {m\over2}\int_{-\infty}^\infty
e^{-\beta}\Big[
{\bf a}^\dagger(\beta){\bf b}(\beta) +
{\bf b}^\dagger(\beta){\bf a}(\beta)\Big] {d\beta\over2\pi}~,
}}
follow directly from \Zzerocharge, \Ycharge\ and the
decompositions \planewaves\ of the fermi fields
$\psi_a$ and~$\psi_b$ in terms of the particle creation and
annihilation operators  ${\bf a}$, ${\bf a}^\dagger$ and ${\bf b}$, 
${\bf b}^\dagger$.

\newsec{Correlation Functions and Particle Matrix Elements}

In this section we use the Ward identities associated with the
$\widehat{SL(2)}$ symmetry~\algebra\ of the doubled model to derive
the differential equation determining the matrix elements of the type 
\bilocal, involving order and/or disorder fields, between any particle 
states. As the warm-up exercise, let us first rederive the celebrated 
equations of Ref.~\McCoy\ for the vacuum-vacuum  matrix elements. 

\subsec{The Correlation Functions}

Consider first the following simple identity
\eqn\ward{
\bra 0|\,\big[\,{\bf 
Z}_2\,,\,\sigma_a(x)\sigma_b(x)~\mu_a(0)\mu_b(0)\,\big]\,| 0\ket = 0~.
}
Evaluating the commutator according to Eq.~\Ztwoaction, one derives
the equation
\eqn\hirotaI{
\partial G\,\partial G - G\,\partial^2 G
- \partial\tilde G\,\partial\tilde G + \tilde G\,\partial^2\tilde G
= 0\,,}
where 
\eqn\Gdef{
G(x) = \bra\,\sigma(x)\sigma(0)\,\ket~, 
\indent 
\tilde G(x) = \bra\,\mu(x)\mu(0)\,\ket~,}
and $\partial = \partial_{\rm z}\,$. Note the signs of the
last two terms, which are due to the presence of the Klein factors in the 
definitions \orderd\ of the fields $\mu_a$, $\mu_b$. The equation similar 
to \hirotaI, with $\partial$ replaced by ${\bar\partial} = 
\partial_{\bar{\rm z}}\,$, is derived 
when one takes ${\bf Z}_{-2}$ instead of ${\bf Z}_{2}$ in \ward. 

Additional equations are derived in similar way. Again, since the
vacuum state $|0 \ket$ is annihilated by all the generators
\Xcharge, \Ycharge, the following equations hold
\eqna\Gward$$\eqalignno{
&\bra 0|\,\big[{\bf X}_{-1},\sigma_a(x)\sigma_b(x)\big]\,
\big[{\bf Y}_1,\mu_a(0)\mu_b(0)\big] - 
\big[{\bf Y}_1,\sigma_a(x)\sigma_b(x)\big]\,
\big[{\bf X}_{-1},\mu_a(0)\mu_b(0)\big]\,|0\ket
\cr
&= i\,{m^2\over2}
\bra0|\,\big[{\bf 
Z}_0,\sigma_a(x)\sigma_b(x)\big]\,\mu_a(0)\mu_b(0)\,|0\ket~,
&\Gward a
\cr
&\bra0|\,\big[{\bf X}_{-1},\sigma_a(x)\mu_b(x)\big]\,
\big[{\bf Y}_1,\mu_a(0)\sigma_b(0)\big] - 
\big[{\bf Y}_1,\sigma_a(x)\mu_b(x)\big]\,
\big[{\bf X}_{-1},\mu_a(0)\sigma_b(0)\big]\,|0\ket
\cr
&= i\,{m^2\over2}
\bra0|\,\big[{\bf 
Z}_0,\sigma_a(x)\mu_b(x)\big]\,\mu_a(0)\sigma_b(0)\,|0\ket~,
&\Gward b
}$$
where ${\bf Z}_0$ in the right hand sides comes from the commutator
$[{\bf X}_{-1}, {\bf Y}_{1}]$ (see Eq.~\algebra). Now using Eqs.
\Xaction\ and \Yaction\ one derives
\eqna\Geq$$\eqalignno{
G\,\partial\bar\partial G - \partial G\,\bar\partial G
+ \tilde G\,\partial\bar\partial \tilde G
- \partial\tilde G\,\bar\partial \tilde G &= 0~,
&\Geq a
\cr
G\,\partial\bar\partial\tilde G - \partial G\,\bar\partial\tilde G
+ \tilde G\,\partial\bar\partial G - \bar\partial G\,\partial\tilde G
&= \(m\over2\)^2 G\,\tilde G~.
&\Geq b
}$$
The Eqs.~\hirotaI\ and \Geq {a,b}\ are known as the quadratic form
\refs{\PerkIII, \MW} of the famous differential equations of Wu, McCoy, 
Tracy and Barouch \McCoy. When $G(x)$ and ${\tilde G}(x)$ are written in 
terms of auxiliary functions $\varphi(x)$ and $\chi(x)$ as 
\eqn\param{
m^{-1/4}\,G = e^{\chi/2}\cosh(\varphi/2)~,
\indent
m^{-1/4}\,{\tilde G} = e^{\chi/2}\sinh(\varphi/2)~,
}
the Eqs.~\hirotaI, \Geq {a,b}\ take the sinh-Gordon form,
\eqn\GeqII{\eqalign{
\partial\bar\partial\varphi &= {m^2\over8}\sinh(2\varphi)~,
\cr
\partial\bar\partial\chi &= {m^2\over8}\Big[1-\cosh(2\varphi)\Big]~,
}
\indent
\eqalign{
\partial^2\chi + \(\partial\varphi\)^2 &= 0~,
\cr
\bar\partial^2\chi + \(\bar\partial\varphi\)^2 &= 0~.
}}
The correlation functions \Gdef\ are related to the rotationally-invariant 
solution $\varphi = \varphi(r)$, $\chi = \chi(r)$, where $r = |x|$ is 
the distance between $x$ and the origin; thus, Eqs.~\GeqII\ reduce to
ordinary differential equations with respect to $r$. The solution relevant 
to the problem can be singled out by specifying its $r\to 0$ asymptotic 
behaviour,
\eqn\sinhshort{
\varphi(r) \sim - \ln{m\,r\over 2} - \ln\(-\Omega\) +
O\(r^4\,\Omega^2\)\,, 
\indent 
\chi(r) \sim {1\over 2}\,\ln\(4m\,r\) + \ln\(-\Omega\) + O\(r^2\)\,;}
here and below in the Appendix we use the notation
\eqn\Omegadef{
\Omega = \ln\({e^\gamma\over 8}m\,r\)~,
}
where $\gamma$ is the Euler's constant. This solution has the property 
that $\varphi(r)$ decays at $r\to\infty$, while $\chi(r)$ approaches a
constant $4\ln {\bar s}$. All these properties are elaborated in 
Ref. \McCoy, where more details can be found. We present some useful 
expansions in the Appendix.

\subsec{One-Particle Matrix Elements}

It is straightforward to extend this technique to the matrix elements
of the form~\bilocal\ involving the particle states. 
First, consider the one-particle matrix elements 
\hbox{$\bra\, 0 \mid \,\sigma(x) \mu(x')\,\mid A(\beta)\, \ket$}. It is 
useful to introduce the following notations
\eqn\onepart{\eqalign{ 
\bra0\mid\sigma(x)\mu(x')\mid A(\beta)\ket &= 
E({x+x'};\beta)\,F(x-x';\beta)\,,
\cr \bra 0\mid\mu(x)\sigma(x')\mid A(\beta)\ket &=
E({x+x'};\beta)\,{\tilde F}(x-x';\beta)\,, }} 
where
\eqn\waves{E({x+x'};\beta) = e^{-({\rm y +
y'})\,{m\over2}\cosh\beta + i({\rm x + x'}){m\over2}\sinh\beta}}
are the ``center of mass'' plane waves. The fact that $F$ and ${\tilde F}$ 
depend on the separation $x-x'$ only is a simple consequence of the 
kinematics. Likewise, from the rotational symmetry of the problem, both 
functions depend only on $r$ and the combination $\vartheta =\theta - 
i\beta$, where $(r,\theta)$ are the polar coordinates associated with the 
separation $x-x'$, i.e. \eqn\polar{ {\rm z}- {\rm z}' = r\,e^{i\theta}\,, 
\indent {\bar{\rm z}}- {\bar{\rm z}}' = r\,e^{-i\theta}\,.}
And it follows from what was said in Section 2 about the local properties 
of the order versus the disorder field that $F(x-x';\beta)$ and ${\tilde 
F}(x-x';\beta)$ are double-valued functions of the coordinates, changing 
the sign when $x$ is brought around $x'$. With this understood, the above 
definition implies $F(x;\beta) = \pm {\tilde F}(-x;\beta)$, where the 
sign depends on the way of continuation $x\to -x$.

The differential equations for the functions $F$ and ${\tilde F}$ can
be derived from the identities
\eqn\Fward{\eqalign{
\bra\, 0\mid\!
{\bf Y}_1\,~\sigma_a(x)\sigma_b(x)~\sigma_a(x')\mu_b(x')\mid 
A(\beta)\,\ket
&= 0~,
\cr
\bra\, 0\mid\!{\bf Y}_1~\,
\sigma_a(x)\mu_b(x)~\sigma_a(x')\sigma_b(x')\mid A(\beta)\,\ket
&= 0~.
}}
One uses the commutation relations \Yaction\ to move the operator
${\bf Y}_1$ to the right, and then applies \YactionII\ to evaluate the
action of this operator on the particle state. The resulting equations
are linear in $F$ and ${\tilde F}$, and involve the correlation
functions \Gdef\ as the coefficients. They are brought to a nice form
by introducing the functions $\Psi_{+}$ and $\Psi_{-}$ related to $F$
and ${\tilde F}$ as
\eqn\paramII{\eqalign{
F+i\,{\tilde F}&= {\bar\omega}\,m^{1/4}\,e^{\chi/2}\,\Psi_{+}~,
\cr
F-i\,{\tilde F}&= {\omega}\,m^{1/4}\,e^{\chi/2}\,\Psi_{-}~,
}}
where again $\omega=e^{i\pi/4}$, $\bar\omega=e^{-i\pi/4}$, and
$\chi$ is related to the two-point correlation functions as in
\param. With these notations, the Ward identities \Fward\ yield
\eqn\FeqI{\eqalign{
\partial\Psi_+ &= -\h\partial\varphi\,\Psi_+ 
+ {m\over4}e^\beta e^\varphi\,\Psi_-~,
\cr
\partial\Psi_- &= \h\partial\varphi\,\Psi_- 
- {m\over4}e^\beta e^{-\varphi}\,\Psi_+~,
}}
where $\partial=\partial_{\rm z}$ and the function $\varphi$ is the
same as in \param. Replacing ${\bf Y}_1$ in \Fward\ with ${\bf 
Y}_{-1}$ one derives similar equations involving the derivative 
${\bar\partial} = \partial_{\bar{\rm z}}\,$, 
\eqn\FeqII{\eqalign{
\bar\partial\Psi_+ &= \h\bar\partial\varphi\,\Psi_+
- {m\over4}e^{-\beta} e^{-\varphi}\,\Psi_-~,
\cr
\bar\partial\Psi_- &= -\h\bar\partial\varphi\,\Psi_-
+ {m\over4}e^{-\beta} e^{\varphi}\,\Psi_+~.
}}

One recognizes in \FeqI, \FeqII\ the Lax representation of the
sinh-Gordon equation: the first of the Eqs.~\GeqII\ guarantees
compatibility of \FeqI\ and \FeqII. Thus, the one-particle matrix
elements \onepart\ are related to the solution of the linear problem
\FeqI, \FeqII\ associated with the sinh-Gordon equation \GeqII.

Let us briefly describe here some elementary properties of the functions
$\Psi_\pm$. As was mentioned above, they depend on two
variables, $r$ and $\vartheta = \theta - i\beta$, where $(r,\theta)$ are
the polar coordinates defined in \polar. The monodromy
properties
\eqn\psimonod{
\Psi_{+}(r,\vartheta+\pi) = i\,\Psi_{+}(r,\vartheta)\,,
\indent \Psi_{-}(r,\vartheta+\pi) = -i\,\Psi_{-}(r,\vartheta)\,,
}
follow directly from the definitions \onepart, \paramII. Also, it is
possible to show that
\eqn\psicross{
\Psi_{+}(r,\vartheta) = \Psi_{-}(r,-\vartheta)\,,}
and that when both $r$ and $\vartheta$ are real (i.e. the rapidity
$\beta$ is continued to pure imaginary values) $\Psi_{+}$ and
$\Psi_{-}$ take complex conjugate values. The $r\to 0$ behaviour
of $\Psi_{+}$ and $\Psi_{-}$ follows from the short-distance operator
product expansion \spinmu, 
\eqn\psishort{e^{\chi(r)/2}\,\Psi_\pm(r,\vartheta) \sim
\sqrt{2\pi}\,(m\,r)^{1/4}\,e^{\pm i\vartheta/2}\,.
}
More details on the asymptotic behaviour is presented in the Appendix.
Note that in view of \psimonod, \psicross, $\Psi_+(r,\vartheta)$ and  
$\Psi_-(r,\vartheta)$ admit the Fourier decompositions of the form
\eqn\psifourier{ \Psi_\pm(r,\vartheta) =
\sum_{n=-\infty}^{\infty}\,\Psi_{2n}(r)\,e^{\pm i\({1\over
2}+2n\)\vartheta}\,, 
}
where $\Psi_{2n}(r)$ are real at real $r$.
The coefficients $\Psi_{2n} (r)$ can be interpreted in terms of
the finite-distance version of the operator product expansion
\spinmu\ , i.e. \eqn\psiope{\sigma(x) \mu(x') =
\sum_{n=0}^{\infty}\,\bigg[C_n (x-x')\,\partial^n
\psi\bigg({{x+x'}\over 2}\bigg) + {\bar C}_n
(x-x')\,{\bar\partial}^n {\bar\psi}\bigg({{x+x'}\over
2}\bigg)\bigg] + \cdots\,,} where the dots stand for the
contributions of the multi-fermion operators (i.e. the composite
fields built from three or more fermi fields, defined in such a
way that their matrix elements between $\bra 0 \mid $ and $\mid
A(\beta) \ket$ vanish). Using the polar coordinates \polar\ to
represent the separation $x-x'$, we have
\eqn\opeCn{C_{n}(r,\theta) = \omega\,e^{i\({1\over
2}+n\)\theta}\,c_n(r)\,,\indent {\bar C}_n (r,\theta) =
{\bar\omega}\,e^{-i\({1\over 2}+n\)\theta}\,c_n (r)\,,} where
\eqn\opecn{\eqalign{ c_{2n}(r)\ \  &= \ \ \ {{(-)^n\,2^{2n-1}\ \
}\over{\sqrt{\pi}\,m^{2n+1/4}}}\,e^{\chi(r)/ 2}~\,\Psi_{2n}(r)\,, 
\cr c_{2n-1}(r) &= -
{{(-)^n\,2^{2n-2}}\over{\sqrt{\pi}\,m^{2n-3/4}}}\,
e^{\chi(r)/2}~\Psi_{-2n}(r)\,.}}

\subsec{Multi-Particle Matrix Elements}

The matrix elements involving more then one particle can be 
determined in a similar way. Here we explicitly elaborate the matrix
elements
\eqn\Gmulti{\eqalign{
G(x,x';\beta_1,\ldots,\beta_{2N}) &=
\bra\,0\mid\sigma(x)\sigma(x')\mid A(\beta_1)\ldots A(\beta_{2N})\,\ket~,
\cr
\tilde G(x,x';\beta_1,\ldots,\beta_{2N}) &=
\bra\,0\mid\mu(x)\mu(x')\mid A(\beta_1)\ldots A(\beta_{2N})\,\ket~.
}}
In order to avoid lengthy expressions, we shall drop the explicit
indication of position dependence from now on and use instead the
notations $G(\beta_1,\ldots,\beta_{2N})$ and $\tilde G(\beta_1, \ldots,
\beta_{2N})$ for the matrix elements \Gmulti. Similarly,
$\Psi_\pm(\beta)$ will stand for the functions defined in \paramII\ in 
relation to the one-particle matrix elements \onepart.

Let us start with the derivation of the two-particle matrix 
elements $G(\beta_1,\beta_2)$ and ${\tilde G}(\beta_1, \beta_2)$. 
It suffices to consider the identities
\eqn\twopartward{\eqalign{
&\bra\, 0\mid\!{\bf Z}_0~\,\sigma_a(x)\mu_b(x)~\sigma_a(x')\mu_b(x')
\mid A(\beta_1)B(\beta_2)\,\ket = 0~,
\cr
&\bra\, 0\mid\!{\bf Y}_1~\,\sigma_a(x)\sigma_b(x)~\sigma_a(x')\sigma_b(x')
\mid A(\beta_1)B(\beta_2)\,\ket = 0~,
\cr
&\bra\, 0\mid\!{\bf Y}_1~\,\mu_a(x)\mu_b(x)~\mu_a(x')\mu_b(x')
\mid A(\beta_1)B(\beta_2)\,\ket = 0~.
}}
Again, one uses the commutators \Zaction, \Yaction, and then
Eqs.~\ZactionII, \YactionII\ to evaluate the action of the operators ${\bf 
Z}_0$ and ${\bf Y}_1$ on the two-particle states. The relations that come 
out allow one to find explicit 
expression for the desired two-particle matrix elements in terms of the 
one-particle ones. The result is
\eqn\twopartII{\eqalign{
i\,{G_+(\beta_1,\beta_2)\over G+\tilde G}
& = {E(\beta_1)E(\beta_2)\over e^{\beta_1} + e^{\beta_2}}
\Big[e^{\beta_1}\Psi_-(\beta_1)\Psi_+(\beta_2)
- e^{\beta_2}\Psi_+(\beta_1)\Psi_-(\beta_2)\Big]~,
\cr
i\,{G_-(\beta_1,\beta_2)\over G-\tilde G}
& = {E(\beta_1)E(\beta_2)\over e^{\beta_1} + e^{\beta_2}}
\Big[e^{\beta_1}\Psi_+(\beta_1)\Psi_-(\beta_2)
- e^{\beta_2}\Psi_-(\beta_1)\Psi_+(\beta_2)\Big]~,
}}
where
\eqn\Gpm{
G_\pm(\beta_1,\beta_2) = G(\beta_1,\beta_2) \pm \tilde G(\beta_1,\beta_2)~,
}
and $E(\beta)\equiv E(x+x';\beta)$ are the plane-wave factors \waves.

Once the two-particle matrix elements are found, all the multiparticle
ones are obtained through the recursive equation
\eqn\Gmultieq{
G\,G(\beta_1,\ldots,\beta_{2N})
= \sum_{i=1}^{2N-1}
(-)^{i+1} G(\beta_1,\ldots,\beta_{i-1},\beta_{i+1},\ldots,\beta_{2N-1})
G(\beta_i,\beta_{2N})~,
}
which follows directly from the Ward identity
\eqn\Gmultiward{
\bra\,0\mid\!{\bf Z}_0~\,\sigma_a(x)\sigma_b(x)~\sigma_a(x')\sigma_b(x')
\mid A(\beta_1)\ldots A(\beta_{2N-1})B(\beta_{2N})\,\ket= 0~.
}
Of course the recursive equation \Gmultieq\ is nothing but the
statement that the multiparticle elements \Gmulti\ are expressed in
terms of the two-particle ones, through all possible Wick pairings,
\eqn\Gmultisol{
G^{N-1}\,G(\beta_1,\ldots,\beta_{2N})
= {1\over(2N)!!}\,\epsilon^{a_1\ldots a_{2N}}
G(\beta_{a_1},\beta_{a_2})\ldots
G(\beta_{a_{2N-1}},\beta_{a_{2N}})~,
}
where $\epsilon^{a_1\ldots a_{2N}}$ denotes usual antisymmetric tensor and
summation over $a_i = 1,\ldots, 2N$ is implicit. This form reflects the 
free-fermion structure of the Ising field theory. The matrix elements 
${\tilde G}(\beta_1,\ldots, \beta_{2N})$ have the same form in terms of 
the functions ${\tilde G}(\beta_1, \beta_2)$.

\newsec{Corrections to the Masses}

As was already mentioned in the Introduction, the primary motivation
of this study was the development of more efficient perturbation
theory for the Ising field theory in the presence of external field. 
The last theory corresponds to adding the spin term to the free action
\action, 
\eqn\perturbation{
{\cal A} = {\cal A}_{\rm FF} + h \int\,d^2 x\,\sigma(x)\,,
}
and it appears to be highly nontrivial interacting field theory.
Its physics depends on a single scaling parameter $\eta =
m/|h|^{8/15}$ and shows rich behavior in terms of its particle
spectrum \mccoywuI. Exact solution exists only in the limiting cases 
$\eta = \pm\infty$, where of course the theory reduces back to the
free fermions \action, and at $\eta = 0$, where it is integrable, 
but at generic values of this parameter the mass spectrum is understood
mostly on a qualitative level, as described in \refs{\mccoywuI, 
\Mussardo, \us}. In this situation, perturbative expansions around $\eta = 
\pm\infty$, say, are useful. Here we will display the calculations of the 
leading perturbative correction, $\sim h^2$, to the particle mass in the
disordered \hbox{regime $\eta \to -\infty$}, and of the same order 
correction to the ``quark mass'' in the ordered regime $\eta\to +\infty$.

Both these calculations are based on the formula \selfenergy, but of
course their interpretations in terms of the physical mass spectrum 
are rather different. In the disordered regime $\eta \to -\infty$ the
spectrum contains a single particle, and the calculation of the mass
correction according to \selfenergy\ yields just that - the correction
to its mass $M_1$, i.e. the coefficient $a$ in
\eqn\highM{
M_1 (\eta) = |m| + a\,{{|m|}\over{(-\eta)^{15/4}}} +
O\big((-\eta)^{-15/2}\big)~, \indent \eta \to -\infty~, 
}
where $m$ is the mass parameter of the unperturbed theory
\action. Note that we put $|m|$ here because in our convention
\asigmaII\ the disordered regime corresponds to negative $m$ in
\action. Recall that all equations in the previous Sections are
written down under the opposite assumption $m>0$. But thanks to the
duality $m \leftrightarrow -m$, $\sigma \leftrightarrow \mu$ we can
still use them in this computation, just taking in \selfenergy\ the 
matrix elements of the disorder fields $\mu$ instead of
$\sigma$. Therefore we have for the coefficient $a$ in \highM, 
\eqn\highcoeff{
a = - {1\over 2}\,|m|^{15/4}\,\int\,\bra\, \beta \mid
\mu(x)\mu(0) \mid \beta \,\ket_{\rm irred} \,d^2 x~.}

On the contrary, in the ordered regime $\eta\to+\infty$ the mass spectrum 
is rather complex. The last term in \perturbation\ creates a long-range
confining iteraction between the original particles of \action, which
become ``quarks''. The particle spectrum consists of a tower of their
meson-like bound states. The meson masses $M_i\,$, $i=1,2,\ldots\,$, do 
not 
admit straightforward perturbative expansions in $h^2$, instead they expand 
in fractional powers of $h$ \refs{\mccoywuI, \us},
\eqn\lowM{
M_i (\eta) = 2\,m_q (\eta) \,\left\{1 +
{{\big(2\,{\bar s}\big)^{2/3}\,z_i}\over{2\,\eta^{5/4}}} - 
{{\big(2\,{\bar s}\big)^{4/3}\,z_{i}^2}\over{40\,\eta^{5/2}}} +
\bigg( {{11\,z_{i}^3}\over 2800} - {57\over 560}\bigg)\,
{{\big(2\,{\bar s}\big)^{2}}\over{\eta^{15/4}}} + \ldots\right\}\,,}
where $\bar s$ relates to the spontaneous magnetization at zero $h$ as
in Eq.~\order, and $-z_i$ are consecutive zeros of the Airy
function, ${\rm Ai}(-z_i) = 0$. Here $m_q$ denotes the ``quark mass'' 
which admits
usual perturbative expansion in powers of $h^2$,
\eqn\quarkM{
m_q (\eta) = m + a_q\,{{m}\over{\eta^{15/4}}}+
O\big(\eta^{15/2}\big)~, \indent \eta \to +\infty~, 
}
where the coefficient $a_q$ is computed according to the Eq.\selfenergy~,
\eqn\lowcoeff{
a_q = - {1\over 2}\,m^{15/4}\,\int\,\bra\, \beta \mid
\sigma(x)\sigma(0) \mid \beta\, \ket_{\rm irred} \,d^2 x~.
}
Note that the leading radiative correction in \quarkM\ competes with 
the last term in \lowM. We also would like to stress here that, since
the ``quarks'' do not appear in the asymptotic states of the theory,
there is no definition of the ``quark mass'' independent of the
perturbation theory. Therefore, while \highcoeff\ is derivable as
usual from the LSZ theory, the perturbative expansion \quarkM\ is
rather a definition of the quark mass. Nonetheless, the leading
correction in \quarkM\ shows up in many places besides the expansion
\lowM\ of the meson masses, notably in the $\eta\to\infty$ expansion
of the ``false vacuum'' resonance decay rate (see \us).

With all this said, we just proceed with evaluating the
integrals in \highcoeff\ and \lowcoeff. The matrix elements involved
are obtained from the matrix elements in Section 4 through usual
crossing equations. One writes the full matrix elements as
\eqna\Gbetabeta$$\eqalignno{
\bra \,A(\beta_1)\mid\sigma(x)\sigma(x')\mid A(\beta_2)\,\ket &=
2\pi\,\delta(\beta_1-\beta_2)\,G + G(\beta_1|\beta_2)~,
&\Gbetabeta a
\cr
\bra \,A(\beta_1)\mid \mu(x)\mu(x')\mid A(\beta_2)\,\ket &=
2\pi\,\delta(\beta_1-\beta_2)\,{\tilde G} + \tilde G(\beta_1|\beta_2)~,
&\Gbetabeta b
}$$
where the delta-function terms are the disconnected ``direct
propagation'' parts shown schematically in Fig.~1, with $G=G(x,x')$ and
$\tilde G={\tilde G}(x,x')$ being the two-point functions~\Gdef, and the
functions $G(\beta_1|\beta_2)$ and ${\tilde G}(\beta_1|\beta_2)$ (whose
arguments $x, x'$ are also supressed) are related to the two-particle
matrix elements $G(\beta_1,\beta_2)$ and ${\tilde G}(\beta_1,
\beta_2)$, Eq.~\Gmulti, as
\eqn\GbetabetaII{
G(\beta_1|\beta_2) = G(\beta_1 - i\pi,\beta_2)~,
\indent
\tilde G(\beta_1|\beta_2) = \tilde G(\beta_1 - i\pi,\beta_2)~.
}

\fig{1}{Disconnected ``direct propagation'' part in \Gbetabeta a; the 
double line represents two-point  correlation function $G(x,x')$ and the 
plain line indicates a particle of rapidity $\beta_1$.} 
{\epsfysize=2.7cm\epsfbox{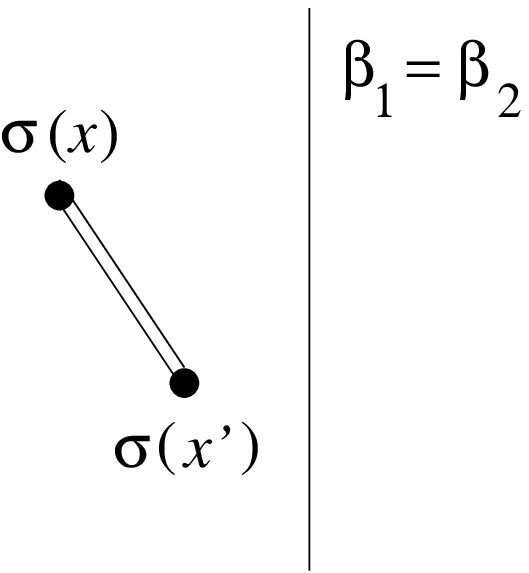}}

The Eqs.~\highcoeff\ and \lowcoeff\ involve only the diagonal elements
${\tilde G}(\beta|\beta)$ and $G(\beta|\beta)$. From~\twopartII\ we
find
\eqn\Gbeta{\eqalign{
G(\beta|\beta) &= \Psi_+(\beta)\Psi_-(\beta)
\[\tilde G - G\,{d\over d\,\beta}\ln\(\Psi_+(\beta)\over\Psi_-(\beta)\)\]~,
\cr
\tilde G(\beta|\beta) &= \Psi_+(\beta)\Psi_-(\beta)
\[G - \tilde G\,{d\over d\,\beta}\ln\(\Psi_+(\beta)\over\Psi_-(\beta)\)\]~.
}}
Remaining disconnected and one-particle reducible parts still have to be 
subtracted before one plugs these functions in \highcoeff, \lowcoeff,
\eqn\subtrac{\eqalign{
\bra\,A(\beta) \mid \mu(x)\mu(x')\mid A(\beta)\,\ket_{\rm irred} = {\tilde
G}(\beta|\beta) - {\tilde S}(\beta|\beta)~,
\cr
\bra\,A(\beta) \mid \sigma(x)\sigma(x')\mid A(\beta)\,\ket_{\rm irred} = 
{G}(\beta|\beta) - {S}(\beta|\beta)~.
}}
Since the matrix elements $\bra A(\beta) \mid \mu (x) \mid A(\beta') \ket$
vanish, but 
\eqn\muone{\bra\, 0 \mid \mu(x) \mid A(\beta)\,\ket =
{\bar\sigma}
\,e^{-{\rm y}\,m\cosh\beta + i{\rm x}\,m\sinh\beta}\,,}
the matrix element of $\mu(x)\mu(x')$ has no one-particle reducible 
parts, and one just has to subtract the disconnected pieces depicted in 
Fig.~2,
\eqn\musub{\eqalign{
{\tilde S}(\beta|\beta) &= 
  \bra A(\beta)|~\mu(x)~|0\ket 
~\bra 0|~\mu(x')~| A(\beta)\ket 
+ \bra 0|~\mu(x)~| A(\beta)\ket
~\bra A(\beta)|~\mu(x')~| 0\ket 
\cr
&= 2\,{\bar\sigma}^2\,\cosh\big(m\,r\sin\vartheta\big)~, 
}}
where $\vartheta = \theta - i\beta$ and $(r,\theta)$ are polar 
coordinates \polar\ associated with the separation $x-x'$. On the other 
hand, $\bra\, 0\mid\sigma(x) \mid A(\beta) \,\ket = 0$, so that the matrix 
element of $\sigma(x)\sigma(x')$ does not have such disconnected parts, but 
it has one-particle reducible components shown in~Fig.~3.

\fig{2}{Disconnected terms in the matrix element 
$\bra\,A(\beta)\mid\,\mu(x)\mu(x')\mid A(\beta)\,\ket$. Sum of these 
diagrams, Eq.~\musub, is the subtraction term $\tilde S(\beta|\beta)$ in 
\subtrac.}
{\epsfysize=3.6cm\epsfbox{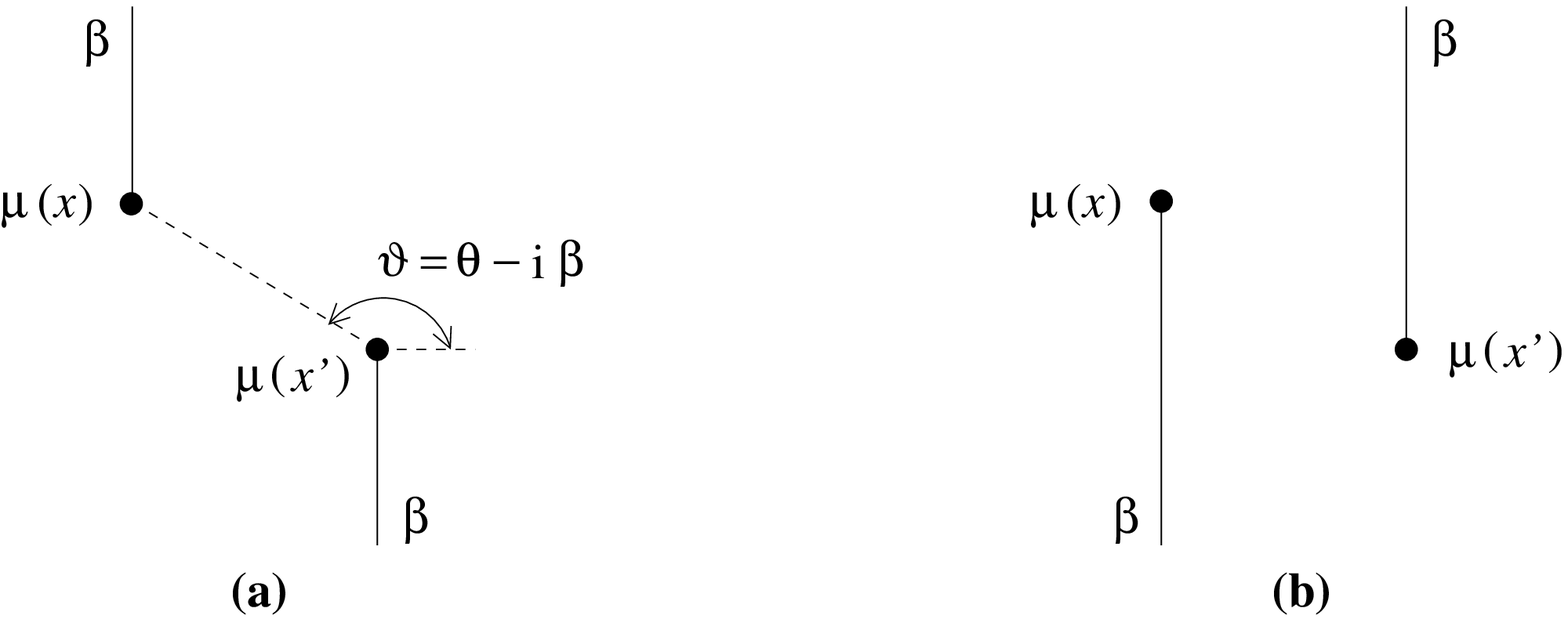}}

\fig{3}{One-particle reducible components in
$\bra\,A(\beta)\mid\,\sigma(x)\sigma(x')\mid A(\beta)\,\ket$. These  add up 
to the subtraction term $S(\beta|\beta)$ in \subtrac\ - see Eq.~\sigmasub.}
{\epsfysize=3.6cm\epsfbox{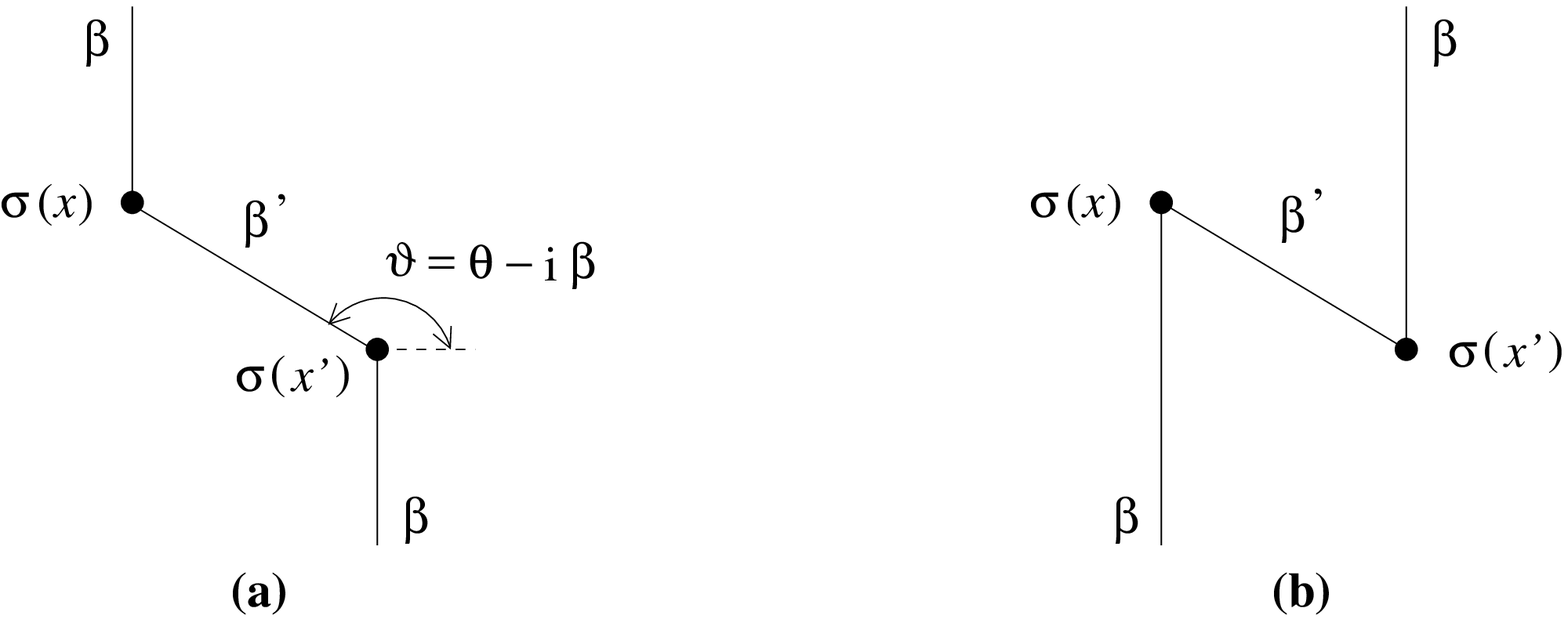}}

This reducible part is written as the principal-value integral 
\eqn\sigmasub{\eqalign{
S(\beta|\beta) =
\int_{-\infty}^{\infty}\,
{{d\beta'}\over {2\pi}}\,
\bigg[&\bra A(\beta)|~\sigma(x)~| A(\beta')\ket~
\bra A(\beta')|~\sigma(x')~|A(\beta)\ket 
\cr
+ \,&\bra 0|~\sigma(x)~|A(\beta)A(\beta')\ket~ 
\bra A(\beta)A(\beta')|~\sigma(x')~|0 \ket\bigg]
}}
over the rapidity $\beta'$ of the intermediate particle in Fig.~3; the 
matrix elements involved here are the $\sigma$-field form-factors 
\Karowski, 
\eqn\formfact{\eqalign{
\bra\, 0\mid \sigma (0) \mid A(\beta_1)A(\beta_2)\,\ket =
i\,\bar\sigma\,\tanh\(\beta_1 -\beta_2\over 2\)\,, 
\cr
\bra\, A(\beta_1)\mid \sigma(0) \mid A(\beta_2)\,\,\ket 
= i\,\bar\sigma\,\coth\(\beta_1 - \beta_2\over 2\)\,,
}}
and in writing \sigmasub\ we have assumed that ${\rm y}\ge{\rm y}'$.
For our purposes the following equivalent form of this term will be more 
convenient,
\eqn\subI{\eqalign{
S(\beta|\beta) = 2\bar\sigma^2\,m\,r\cos\vartheta
+ \bar\sigma^2\, e^{-m\,r\sin\vartheta}\[{1\over\pi}K_0(m\,r) 
+ 2\,\partial_\vartheta A(r,\vartheta-\pi/2)\]&
\cr
+~\bar\sigma^2\, e^{m\,r\sin\vartheta}\[{1\over\pi}K_0(m\,r) 
+ 2\,\partial_\vartheta B(r,\vartheta-\pi/2)\]&~,
}}
where
\eqn\AB{\eqalign{
A(r,\vartheta) &= \int_{-\infty}^\infty{d\beta'\over2\pi i}\,
\tanh\(\beta'-i\vartheta\over2\)e^{-m\,r\cosh\beta'}\,,
\indent -\pi<\Re e\,\vartheta<\pi\,,
\cr
B(r,\vartheta) &= \int_{-\infty}^\infty{d\beta'\over2\pi i}\,
\coth\(\beta'-i\vartheta\over2\)e^{-m\,r\cosh\beta'}\,,
\indent\quad~\, 0<\Re e\,\vartheta<2\pi\,.
}}
Here again $\vartheta = \theta -i\beta$, and $(r,\theta)$ are the polar 
coordinates \polar. The above integrals define the functions 
$A(r,\vartheta)$, $B(r,\vartheta)$ in the specified domains of $\vartheta$ 
only; outside them, they are defined by analytic continuation.	

Using \subtrac, the integrals in \highcoeff\ and \lowcoeff\ were evaluated 
numerically, after finding numerical solution of the differential 
equations \FeqI, \FeqII\ for the functions $\Psi_{\pm}$ in \Gbeta. Since 
the integrals do not depend on $\beta$ we set \eqn\setbeta{\beta=0\,,} 
so that $\vartheta = \theta$, just the euclidean angle between the
separation $x-x'$ and the {\rm x}-axis.
In this case (and indeed at any pure imaginary $\beta$) the functions 
$\Psi_{+}$ and $\Psi_{-}$ are complex-conjugate to each other. 
Few words about integration of the 
differential equations \FeqI, \FeqII\ are worth saying. 
We found it most convenient to use the representation of
the solution in terms of the Backl\"und transformation of the 
angular-symmetric solution $\varphi(r)$, $\chi(r)$ of the sinh-Gordon system
\GeqII\ appearing in the representation \param\ of the correlation
functions. The solution of the linear system \FeqI, \FeqII\ can be
written in terms of $\chi(r)$ and two auxiliary
functions $\phi(r,\vartheta)$ and $\rho(r,\vartheta)$, 
\eqn\paramIII{
e^{\chi/2}\,\Psi_\pm = e^{(\rho \pm i\phi)/2}~,
}
where $\phi$ and $\rho$ must satisfy the equations
\eqn\backlunda{
\partial\big(\varphi + i\phi) 
= {m\over2}\,e^\beta\,\cosh\big(\varphi - i\phi\big)\,, 
\indent
{\bar\partial}\big(\varphi - i\phi) 
= {m\over 2}\,e^{-\beta}\,\cosh\big(\varphi + i\phi\big)\,,}
and
\eqn\backlundb{
\partial\big(\rho - \chi) 
= {m\over 2}\,e^\beta\,\sinh\big(\varphi - i\phi\big)\,, 
\indent
{\bar\partial}\big(\rho-\chi) 
= {m\over 2}\,e^{-\beta}\,\sinh\big(\varphi + i\phi\big)\,,
}
and hence also the sine-Gordon system
\eqn\FeqIII{\eqalign{
\partial\bar\partial\phi &= -{m^2\over8}\sin(2\phi)~,
\cr
\partial\bar\partial\rho &= {m^2\over8}\Big[1+\cos(2\phi)\Big]~,
}
\indent
\eqalign{
\partial^2\rho -\(\partial\phi\)^2 &= 0~,
\cr
\bar\partial^2\rho -\(\bar\partial\phi\)^2 &= 0~.
}}
The solution of \FeqIII\ which appears through \backlunda\ is rather
interesting. The phase $\phi$ is not a single-valued function of the
coordinates. Instead, when written in terms of the polar coordinates
$(r,\vartheta)$, it is quasiperiodic function of the angle, 
$\phi(r,\vartheta+2\pi) = \phi(r,\vartheta) + 2\pi$, as demanded 
by the monodromy properties of the matrix elements \onepart\
stated in Section~4. Qualitatively, it can be described as the
juxtaposition of two sine-Gordon domain walls (i.e. the sine-Gordon
soliton solutions, in the euclidean nomenclature) of oposite sign,
extending along the ${\rm x}$-axis in oposite directions; the solution
is singular at $r=0$, and its shape in the ``junction'' region, 
$r\sim m^{-1}$, is shown in Fig.~4.

\fig{4}{Plot of the phase $\phi(r,\vartheta)$ for fixed $r$ as function 
of $\vartheta = \theta$. It satisfies the symmetry properties
$\phi(r,\vartheta) + \phi(r,\pi-\vartheta) = \pi$ and 
$\phi(r,\vartheta) = -\phi(r,-\vartheta)$. The ``domain 
walls'' at $\theta\approx0$ and $\theta\approx\pi$ are clearly visible in 
the plots for $r\ge3m^{-1}$.} 
{\epsfysize=6cm\epsfbox{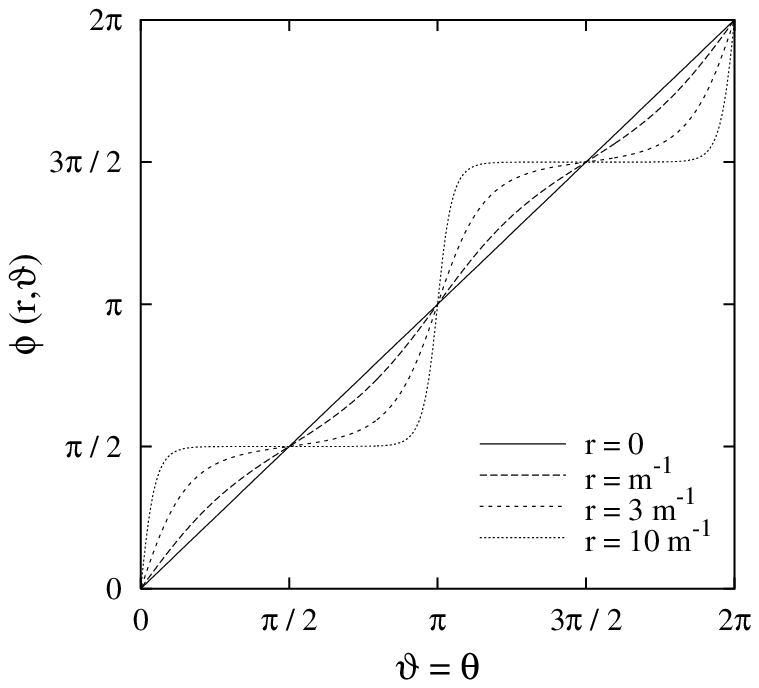}}

The numerical integration was performed in three steps. First, the
ordinary differential equations, the radial form of \GeqII, were
integrated numerically, the initial conditions at small $r$ being provided by 
the short-distance expansions \Gshort, and the large-distance asymptotics 
\Glarge\ were used for precision control. Next, the functions
$\phi(r,\vartheta)$ and $\rho(r,\vartheta)$ were computed by
integrating the first-order differential equations \backlunda,
\backlundb, with the initial conditions fixed using the short-distance
expansions \Fsmall. This step provides the numerics for $G(0|0)$ and
${\tilde G}(0|0)$ as functions of $(r,\vartheta)$; according to \Gbeta,
\eqn\GbetaII{\eqalign{
G(\beta|\beta) &=
e^{\rho-\chi/2}\Big[\sinh(\varphi/2)
- \cosh(\varphi/2)\,\partial_\vartheta\phi\Big]~,
\cr
\tilde G(\beta|\beta) &=
e^{\rho-\chi/2}\Big[\cosh(\varphi/2)
- \sinh(\varphi/2)\,\partial_\vartheta\phi\Big]~.
}}
After the subtractions in \subtrac\ are made to eliminate the reducible 
parts, the integrands in \highcoeff\ and \lowcoeff\ decay exponentially at 
large $r$, rendering the integrals convergent. So, as the last step, the 
integrations over the euclidean coordinates $r,\theta$ were performed 
numerically, yielding
\eqn\finala{a = 10.7619899(1)~,}
and
\eqn\finalb{a_q = {\bar s}^2~0.142021619(1)~.}
These numbers provide more accurate data for the mass corrections previously
estimated in \us.

\newsec{Discussion}

The purpose of this paper was two-fold. Firstly, we have presented
a new derivation of well-known result - the nonlinear differential
equations of Ref.~\McCoy\ for the correlation functions in the Ising
field theory with zero magnetic field. This part has a methodical
value at the best. Nonetheless, we decided to include it here
because we believe our derivation is somewhat simpler then the
traditional ones, and also it has certain potential for
generalizations. It is based on the Ward identities associated
with the special integrals of motion of the doubled Ising field
theory. Incidentally, the ``doubling'', albeit convenient, is not
the most essential part of our approach. The single Ising field
theory has a system of local integrals of motion powerful enough
to render similar derivation possible; the ``doubling'' trick just
makes it shorter. Anyhow, our approach can be adopted to yield
simple derivations of other results in the Ising field theory.
Thus, it was already used in \us\ in deriving the finite-size
form-factors of the spin operator, and recently in \youguys\ in a
simple derivation of the differential equations of \tracy\ in the
Ising field theory on a Poincar\'e disk. Among interesting
potential applications let us mention the Ising field theory at
finite temperature $T$ (equivalently, the theory defined on
euclidean cylinder, with the points $({\rm x}, {\rm y})$ and $({\rm
x},{\rm y}+T^{-1})$ identified). One can notice that all arguments
of Section 4.1 remain valid when the vacuum expectation values
$\bra\, 0 \mid \ldots \mid 0\, \ket$ are replaced by the thermal averages
${\rm tr}\big(e^{-{\bf H}/T} \ldots\big)$, because the total
Hamiltonian ${\bf H}$ of the doubled theory commutes with all the
generators ${\bf Z}_{\pm 2}$, ${\bf X}_{\pm 1}$ and ${\bf Y}_{\pm
1}$ in \ward\ and \Gward {a,b}. It follows that the Eqs. \param\ and
\GeqII\ remain valid in this situation as well - the result
previously obtained by different methods in \PerkI\ and \saleur\
\foot{Additional equation proposed in \saleur, the Eq. $(4.42d)$, 
apparently does not hold.}. In this case the functions $\varphi$ and $\chi$ 
depend on two variables~$({\rm x}, {\rm y})$, but the relevant solution of 
the Eqs. \GeqII\ still can be fixed uniquely by imposing the Matsubara
boundary conditions $\varphi({\rm x}, {\rm y}+T^{-1}) =
\varphi({\rm x}, {\rm y})$, as well as appropriate asymptotic
conditions at $|{\rm x}| \to \infty$ and $({\rm x},{\rm y}) \to
(0,0)$. We plan to exploit advantages of our technique in relation to this 
problem elsewhere. Yet another potential application is in the Ising field
theory with boundaries. On the other hand, extension of this
approach to other integrable field theories is more problematic.
The fact that all descendants of the spin fields up to
sufficiently high level are expressed through the derivatives, as
in the Eqs. \asigmaIII, \asigmaIV, was very essential for
the whole scheme to work; this fact seems to be rather specific
feature of the Ising field theory.

Secondly, the results of Section 4 were used in Sect.~5 in
computing the leading perturbative mass corrections in the Ising
field theory with magnetic field \perturbation, and from this
point of view the numbers \finala\ and \finalb\ constitute the
main result of this work. This computation is a part of our
ongoing project of systematic study of the field theory
\perturbation. We are planning to further apply the technique
developed here to perturbative calculations of other relevant
quantities, such as scattering amplitudes and various structure
functions.

\vskip 0.3in

\centerline{\bf Acknowledgments}

\vskip 0.1in

P.F. is grateful to B.~Doyon and S.~Lukyanov for helpful discussions and 
thanks S.~Ashok, F.~Lesage, J.--M.~Maillet and J.~Rasmussen for
conversations. A.Z.~acknowledges discussions with S.~Lukyanov, F.~Smirnov 
and Al.~Zamolodchikov. This research was supported by the DOE grant 
\#DE-FG02-96\ ER\ 40959. The work of P.F. was also partially supported by 
the grant POCTI/FNU/38004/2001--FEDER (FCT, Portugal).

\vskip 0.1in

\appendix{A}{}

Here we present the short- and large-distance expansions of the
functions $\varphi(r),\ \chi(r)$ in \param\ (mostly borrowed from
Ref.~\McCoy ), as well as the analogous expansions of the
functions $\Psi_{\pm}(r,\vartheta)$ in \paramII. Here again
$\vartheta = \theta - i\beta$, where $(r,\theta)$ are the polar
coordinates \polar. These expansions are used in the numerical
evaluation of the self-energy parts \highcoeff\ and \lowcoeff, as
explained in Section 5.

\bigskip

\noindent {\bf A.1}

The functions $\phi(r)$, $\chi(r)$ in \param\ obey the radial form
of the Eqs. \GeqII, i.e. \eqn\Geqshort{
\partial_r^2\varphi + \textstyle{1\over r}\partial_r\varphi =
\textstyle{m^2\over2} \sinh(2\varphi)~, \indent {2\over
r}\partial_r\chi = \(\partial_r\varphi\)^2 +
\textstyle{m^2\over2}\(1-\cosh(2\varphi)\)~. } As is explained in
\McCoy , the relevant solution is characterized uniquely by its
short-distance asymptotic behavior \sinhshort. Corrections to
this leading asymptotics can be obtained by iterations of
\Geqshort,
\eqn\Gshort{\eqalign{
\varphi(r) &= -\ln\(m\,r\over2\) - \ln\(-\Omega\)
+ (m\,r)^4\,f_4 + (m\,r)^8\,f_8 + O\((m\,r)^{12}\,\Omega^6\)\,,
\cr
\chi(r) &= \h\ln(4m\,r) + \ln(-\Omega) + {(m\,r)^2\over8}
+ (m\,r)^4\,h_4 + (m\,r)^8\,h_8 + O\((m\,r)^{12}\,\Omega^6\)\,,
}}
where $\Omega$
was  defined in \Omegadef, and the coefficients $f_n$ and $h_n$ are
rational functions of $\Omega$,
\eqn\varphicoeffs{\eqalign{
f_4 &= 
-{1\over2^{11}\,\Omega}\(2\,\Omega-1\)\(4\,\Omega^2-2\,\Omega+1\)\,,
\cr
f_8 &= -{1\over2^{28}\,\Omega^2}\(
2048\,\Omega^6 - 4096\,\Omega^5 + 3648\,\Omega^4 - 1568\,\Omega^3
+ 136\,\Omega^2 + 111\,\Omega - 32\)\,,
}}
and
\eqn\chicoeffs{\eqalign{
h_4 &= -{1\over2^{11}\Omega}\(8\,\Omega^3 - 8\,\Omega^2 +
2\,\Omega+1\)\,,
\cr
h_8 &= -{1\over2^{28}\,\Omega^2}\(
2048\,\Omega^6 - 4096\,\Omega^5 + 3776\,\Omega^4 - 1888\,\Omega^3 +
496\,\Omega^2 - 111\,\Omega+32\)\,.
}}
These equations extend the corresponding results presented in \McCoy.

Alternatively, the same solution can be characterized by its
large-distance behavior~\McCoy,
\eqn\Glarge{\eqalign{
\varphi(r) &= {2\over\pi}K_0(m\,r) + O\(e^{-3m\,r}\)\,,
\cr
\chi(r) &= 4\ln\bar s - {2m\,r\over\pi^2}
\Big[m\,r\[\,K_0^2(m\,r)-K_1^2(m\,r)\]
+ K_0(m\,r)K_1(m\,r)\Big] + O\(e^{-4m\,r}\)\,,
}}
where $K_\nu$ are usual modified Bessel functions and $\bar s$ is defined 
in \order.

\bigskip

\noindent {\bf A.2}

The short-distance expansions of $\Psi_{\pm}$ are best written in
terms of the representation \paramIII. The functions
$\phi(r,\vartheta)$, $\rho(r,\vartheta)$ in \paramIII\ solve the
first-order differential equations \backlunda, \backlundb,
with the leading $r\to 0$ asymptotic $\phi(r,\vartheta) \to
\vartheta$, $\rho(r,\vartheta) \to {1\over 2}\,\ln\(4\pi^2 m\,r\)$
(see Eq.~\psishort). Corrections to this asymptotic are
obtained directly, by iterating \backlunda, \backlundb, using
the expansions \Gshort\ for $\varphi(r)$ and $\chi(r)$. We present
just few first terms:
\eqn\Fsmall{\eqalign{
\phi(r,\vartheta) &= \vartheta + (m\,r)^2\,p_2 + (m\,r)^4\,p_4 
+ (m\,r)^6\,p_6 + O\((m\,r)^8\,\Omega^4\)\,,
\cr
\rho(r,\vartheta) &= \h\ln(4\pi^2 m\,r) + (m\,r)^2\,q_2 + (m\,r)^4\,q_4
+ (m\,r)^6\,q_6 + O\((m\,r)^8\,\Omega^4\)\,,
}}
where again $\Omega$ is the logarithm \Omegadef,
\eqn\phicoeffs{\eqalign{
p_2 &= -{1\over2^3}\,\Omega\,\sin 2\vartheta\,,
\cr
p_4 &= {1\over 2^{10}}\,\Omega\,(4\,\Omega-1)\sin 
4\vartheta\,,
\cr
p_6 &= {1\over 2^{15}\, 3}\[
3\(10\,\Omega^2-7\,\Omega+2\)\sin 2\vartheta
-\Omega\(16\,\Omega^2-6\,\Omega+1\)\sin 6\vartheta\]\,,
}}
and
\eqn\rhocoeffs{\eqalign{
q_2 &=
{1\over8} + {1\over 2^4}\(2\,\Omega-1\)\cos 2\vartheta\,,
\cr
q_4 &= {1\over2^{12}}\[(16\,\Omega-8)-
\(16\,\Omega^2-4\,\Omega+1\)\cos 4\vartheta\]\,,
\cr
q_6 &= {1\over2^{16}\,3^2}\[
-9\(12\,\Omega^2-10\,\Omega+3\)\cos 2\vartheta
+\(96\,\Omega^3-36\,\Omega^2+6\,\Omega-1\)\cos 6\vartheta\]\,.
}}
Large-distance behavior of these functions follows from exact
expansions 
\eqn\exactexpan{\eqalign{
F+i\,\tilde F &= \bar\sigma^2\,\sum_{n=0}^\infty\,{e^{-i\pi n/2}\over 
n!}
\Big[e^{{m\over2}r\sin\vartheta}B_n(r,\vartheta-\pi/2)
+ i\,e^{-{m\over2}r\sin\vartheta}A_n(r,\vartheta-\pi/2)\Big]\,,
\cr
F-i\,\tilde F &= \bar\sigma^2\,\sum_{n=0}^\infty\,{e^{i\pi n/2}\over 
n!}
\Big[e^{{m\over2}r\sin\vartheta}B_n(r,\vartheta-\pi/2)
- i\,e^{-{m\over2}r\sin\vartheta}A_n(r,\vartheta-\pi/2)\Big]\,,
}}
of the matrix elements \onepart. When $\vartheta$ lays inside the
strip $0 < \Re e\, \vartheta < \pi$, the functions $A_n
(r,\vartheta)$ and $B_n (r, \vartheta)$ here are defined as the
$n$-fold integrals
\eqn\AnBn{\eqalign{
A_n(r,\vartheta) &= \prod_{1\le j\le n}\bigg[
\int_{-\infty}^\infty{d\beta_j\over2\pi i}
\tanh\(\beta_j-i\vartheta\over2\)e^{-m\,r\cosh\beta_j}\bigg]
\prod_{1\le j < k\le n}\tanh^2\(\beta_j-\beta_k\over2\)\,,
\cr
B_n(r,\vartheta) &= \prod_{1\le j\le n}\bigg[
\int_{-\infty}^\infty{d\beta_j\over2\pi i}
\coth\(\beta_j-i\vartheta\over2\)e^{-m\,r\cosh\beta_j}\bigg]
\prod_{1\le j < k\le n}\tanh^2\(\beta_j-\beta_k\over2\)\,,
}}
($A_0(r,\vartheta) = B_0(r, \vartheta) = 1$) and analytic continuation of 
these functions to full complex $\vartheta$-plane is achieved using the 
following relations
\eqn\ABspropI{\eqalign{
A_n(r,-\vartheta) &= (-1)^n A_n(r,\vartheta)\,,
\cr
B_n(r,-\vartheta) &= (-1)^n B_n(r,\vartheta)
- 2n(-1)^n\,e^{-r\cos\vartheta}A_{n-1}(r,\vartheta)\,,
}}
and
\eqn\ABspropII{\eqalign{
A_n(r,\vartheta+\pi) &= B_n(r,\vartheta)
- 2 n\, e^{-r\cos\vartheta}A_{n-1}(r,\vartheta)\,,
\cr
B_n(r,\vartheta+\pi) &= A_n(r,\vartheta)\,.
}}
The last relations ensure right monodromy properties of the matrix
elements \onepart. The expansions \exactexpan\ are easily obtained in a
standard way, by writing down the intermediate-state decompositions of
the matrix elements \onepart, and using the exact form-factors of
\Karowski. In fact, the series representations \exactexpan\ can be
taken as the starting point of alternative derivation of the equations
\FeqI, \FeqII\ \foot{This was done independently by F.~Smirnov
(private communication).}.

\listrefs

\bye